# Strain-sensitive flexible magnetoelectric ceramic nanocomposites

**Authors**

Minsoo Kim[1]†, Donghoon Kim[1]†, Buse Aktas[1], Hongsoo Choi[2], Josep Puigmartí-Luis[3,4], Bradley J. Nelson[1], Xiang-Zhong Chen[1]*, Salvador Pané[1]*

**Affiliations**

[1]Multi-Scale Robotics Lab, Institute of Robotics and Intelligence Systems, ETH Zurich; Tannenstrasse 3, 8092 Zurich, Switzerland

[2]Department of Robotics & Mechatronics Engineering, DGIST-ETH Microrobotics Research Center; Daegu Gyeong-buk Institute of Science and Technology (DGIST), Daegu, Republic of Korea

[3]Department of Physical Chemistry, Institut de Química Teòrica i Computacional, University of Barcelona; Martí i Franquès, 1, 08028, Barcelona, Spain

[4]Institució Catalana de Recerca i Estudis Avançats (ICREA); Pg. Lluís Companys 23, Barcelona 08010, Spain

\* Corresponding authors. Email: chenxian@ethz.ch, vidalp@ethz.ch
† These authors contributed equally to this work

**Abstract**

Advanced flexible electronics and soft robotics require the development and implementation of flexible functional materials. Magnetoelectric (ME) oxide materials can convert magnetic input into electric output and vice versa, making them excellent candidates for advanced sensing, actuating, data storage, and communication. However, their application has been limited to rigid devices due to their brittle nature. Here, we report flexible ME oxide composite ($BaTiO_3$/$CoFe_2O_4$) thin film nanostructures that can be transferred onto a stretchable substrate such as polydimethylsiloxane (PDMS). In contrast to rigid bulk counterparts, these ceramic nanostructures display a flexible behavior and exhibit reversibly tunable ME coupling via mechanical stretching. We believe our study can open up new avenues for integrating ceramic ME composites into flexible electronics and soft robotic devices.

**Teaser**

We report flexible magnetoelectric oxide nanostructures with large elastic strain and tunable magnetoelectric coupling.

**MAIN TEXT**

**Introduction**

Novel functional materials can provide innovations in flexible electronics and soft robotics for advanced sensing, actuation, and communication applications (*1-4*). While polymer-based materials are considered promising for flexible devices due to their elastic properties, their functional performance is still inferior to those of inorganic materials, such as ceramics. For example, commercially available ceramics possess piezoelectric constants that are orders of magnitude higher than those of piezoelectric polymers (e.g. $BaTiO_3$: 191 pC/N vs. P(VDF-TrFE): 25-30 pC/N) (*5, 6*). Although the rigid mechanical nature of ceramic materials has hindered their utilization in flexible devices, recent observations of abnormal mechanical properties in nanoscale ceramic materials could resolve these issues. Specifically, free-standing nanoscale oxide thin films exhibit distinguished mechanical properties such as superelasticity (*7-9*) and extreme tensile strain endurance (*10, 11*), in contrast to brittle bulk ceramics or thin films on substrates. Additionally, the free-standing thin films can be transferred onto flexible or stretchable substrates (*12-14*), which highly facilitates the integration of these functional oxides into flexible electronics and soft robotics.

Functional ceramic oxides that show coupling between mechanical, electrical, and magnetic order parameters (*15*) are appealing for a wide range of applications in electronics. Magnetoelectric (ME) composites composed of piezoelectric and magnetostrictive oxides are a clear example of this class of materials, where the coupling is achieved through interfacial elastic interactions. As ME composite oxides can detect magnetic fields and convert them into electric outputs, they have been widely applied in smart sensors (*16, 17*), memories (*18, 19*), and biomedical systems (*20*). Extending these superior ME devices to flexible systems would require strong ME coupling and excellent elastic mechanical properties. However, these highly functional ME composite oxides have not been incorporated into flexible systems due to their hard and brittle nature in bulk.

Here, we realize flexible ME composite oxides by creating $BaTiO_3/CoFe_2O_4$ (BTO/CFO) free-standing epitaxial bilayer thin films. We show these free-standing thin films have large elastic strain (> 4 %) and can be transferred to flexible substrates, such as polydimethylsiloxane (PDMS), even with complex geometries. We also investigate the ME performance of these composite oxides. The free-standing and transferred BTO/CFO structures exhibit an order of magnitude larger ME coupling compared to thin films on a substrate as they are free from the substrate clamping (*21*). We also investigate how applied deformations influence the ME coupling and determine the strain dependency of the ME coupling in the transferred BTO/CFO structures by stretching the PDMS substrate. Our study not only represents a strategy for implementing ME oxide ceramics in flexible electronics and soft robotic devices, but also provides insight into their performance.

**Results**

Fig. 1 shows the flexible ME composites transferred onto a PDMS substrate that can endure bending and stretching. The free-standing BTO/CFO structures were fabricated using photolithography, dry etching, and wet etching processes sequentially (Fig. 1a). Different patterns were designed on the as-deposited BTO/CFO films through photolithography and Ar-ion milling. The patterned nanostructures were then released through selective etching of the underlying MgO substrate followed by critical point drying (details in Materials and Methods).

These free-standing BTO/CFO structures were transferred to a stretchable PDMS substrate by direct stamping. The transferred structures showed good flexibility, stretchability, and endurance to different types of deformations, for example, bending through the PDMS substrate (Fig. 1b).

We quantitatively investigated the stretchability of BTO/CFO free-standing films by nanomechanical tensile test. In contrast to bulk ceramics, extreme elongations were observed in the dogbone-like structures (Fig. 2a, Movie S1, details in Materials and Methods) (*22*). The stress-strain curve from the fracture test shows maximum failure stress and strain of 1514 MPa and 3.26 % (blue, Fig. 2b), respectively. The tensile strain is significantly larger than that of bulk ceramic materials. Although earlier fractures have been observed, including failure stress and strain of 0.89 % and 301.2 MPa (grey) and 1.81 % and 772.6 MPa (red), the slope of each stress-strain curve, namely, the Young's modulus ($E = \frac{\sigma}{\varepsilon}$) of the structure, is almost identical for all three cases with a value of 48.8 GPa (± 1.87 GPa), which indicates the reliability of the measurements and the repeatability of the performance of the material. The early fracture points can be ascribed to (i) stress concentration near the markers and (ii) uneven contact between the structure and the force sensor. In most cases, fractures occurred near the marker or the contact point between the sample and the force sensor (Fig. S1), indicating an early fracture caused by the stress concentration. Mechanical analysis using the finite-element method further corroborates the hypothesis that stress is concentrated near the marker points and the contact point (Fig. 2c). While the exact strain analysis was difficult to ascertain without markers, fractures occurred in the middle of the neck in some cases (Fig. S2). In addition, AFM topography images (Fig. S3) of the stretched BTO/CFO (a tensile strain of ~ 4.1 %) without displaying any cracks further indicate a larger failure tensile strain in BTO/CFO free-standing structures. These enhanced fracture stress and strain values can enable ME composites to be used in flexible applications.

Interestingly, the Young's modulus of BTO/CFO free-standing films is much lower than that of bulk BTO (~120 GPa) and bulk CFO (~188 GPa) (*23-26*). Ceramic materials that are brittle and tough in bulk can show abnormal mechanical behavior at the nanoscale, such as ultrahigh- or superelasticity and low Young's modulus (*7, 27, 28*). Unlike bulk materials, surface and interface contributions can be accounted for in the unique tensile mechanical properties of the nanoscale BTO/CFO free-standing structures. In particular, surface stress, originating from surface relaxation and reconstruction, and interface stress, originating from the lattice mismatch, can result in a softening of the elastic moduli and an increase in the maximum elastic strain (*29, 30*). In addition, the scarcity of internal defects in nanoscale structures compared to bulk materials further restrict dislocation multiplication and crack evolution under tensile stress, resulting in an increase in the elastic limit (*31-33*).

We then investigated the ME coupling characteristics in transferred BTO/CFO structures on a PDMS substrate (Fig. 3a). Transferred BTO/CFO structures showed excellent ME coupling, as shown in Fig. 3b and 3c. Coercive field shifts in piezoelectric force microscopy (PFM) hysteresis loop measurements have been clearly observed with the application of magnetic fields and higher magnetic fields resulted in higher coercive field shifts. As their electrical and magnetic properties highly depend on strain, we also tested their ME coupling behavior while stretching the PDMS substrate. For quantitative analysis, we define the ME coefficient as $\alpha_E = \Delta E / \Delta H$, where $\Delta H$ is the applied in-plane magnetic field and $\Delta E$ is the measured out-of-plane electric field that can be estimated by the shift of the center of the hysteresis loop (*34*). Fig. 3d shows ME coupling coefficients as a function of external tensile strain. Although the coercive field shifts and the ME coupling coefficient decreased with larger tensile strain, ME coupling was observed with up to 4.1 % of tensile strain. As the applied tensile strain increased from 0 % to 4.1 %, the coupling coefficient decreased from $16.6 \times 10^5$ mV cm$^{-1}$ Oe$^{-1}$ to $2.04 \times 10^5$ mV cm$^{-1}$ Oe$^{-1}$. In addition, we also tested the repeatability and recoverability of the ME coupling change in transferred BTO/CFO structures by stretching and releasing the PDMS substrate repeatedly. ME coupling coefficients were measured *in situ* with the strain range from 0 % to 3 % during stretch-release cycling of the PDMS substrate. As can be seen in Fig. 3e, the ME coupling coefficient decreased during stretching, but recovered its values during releasing. This behavior was observed for several cycles.

To analyze the effect of the geometrical configuration on ME coupling behavior, we compared the ME coupling coefficients in BTO/CFO structures of different mechanical boundary conditions, including free-standing structures, stretchable structures on a PDMS substrate, square-disk patterned structures on a rigid substrate, and as-deposited thin films (Fig. 4). Higher mechanical degrees of freedom in ME composites results in a higher ME coupling coefficient, which can be attributed to the reduced substrate clamping in the BTO/CFO (*21*). Adopting different fabrication techniques, a wide range of ME coupling coefficients could be obtained with the same BTO/CFO epitaxial bilayer thin films. In particular, in BTO/CFO stretchable structures on PDMS, real-time tuning of ME coupling strength is possible, in contrast to the patterned thin films on substrates where the ME couplings are fixed once films have been fabricated (*21*).

The strain-dependent ME coupling in BTO/CFO on the PDMS substrate can be attributed to three possible factors: (i) tensile strain from the PDMS substrate exerts a 'clamping' effect on the CFO layer, preventing effective magnetostriction and interfacial interaction with BTO, and eventually resulting in weak ME coupling. (ii) the magnetostriction of CFO also depends on the external mechanical stress. In a simplified magneto-elastic model which considers unidirectional stress in magnetic materials, the magnetostriction coefficient can be expressed as

$$\lambda = \lambda_{100}\left(1 - \frac{3\left(exp\left(\frac{3}{2}A_s\lambda_{100}\sigma_{yy}\right) + 1\right)}{2\left(cosh(\kappa H) + exp\left(\frac{3}{2}A_s\lambda_{100}\sigma_{yy}\right) + 1\right)}\right)$$

where $A_S = \frac{3\chi^0}{\mu_0 M_S}$, $\kappa = \mu_0 A_s M_s$, $\mu_0$ is the vacuum permittivity, $M_s$ is the saturation magnetization, $\lambda_{100}$ is the magnetostrictive coefficient, and $\chi^0$ is the initial slope of the unstressed anhysteretic magnetization curve (*35, 36*). Therefore, the increase in the uniaxial tensile stress $\sigma_{yy}$ gives rise to a decrease in the magnetostriction $\lambda$ and eventually a weak ME coupling. (iii) under large tensile strain, BTO shows negligible out-of-plane polarization ($P_{op}$) as well as weaker strain sensitivity, $\frac{\partial P_{op}}{\partial \varepsilon}$ (*37-40*). Therefore, diminished ME coupling can be expected under large tensile strain since the magnetostriction of CFO would cause only a small change in $P_{op}$ in BTO.

Our results indicate that ME coupling of transferred BTO/CFO structures can be systematically tuned in real-time by applying different tensile strains via stretching the PDMS substrate, unlike thin films on rigid substrates where external strains are fixed by the lattice parameters of the substrate. The reversibly tunable ME coupling also provides opportunities for BTO/CFO transferred structures to be used for flexible and stretchable magnetoelectronics for strain-sensing applications. In flexible and stretchable magnetoelectronics, the ability to fabricate various shapes and figures on different substrates allows for multiple designs and further expansion of their applications. To this end, the adopted stamp-transfer method is a versatile approach for the realization of BTO/CFO architectures with different sizes and shapes. In Fig. 5, we show various designs transferred by the stamping method, including positively and negatively engraved ETH alphabet logos and connected-rectangular arrays. From just a few to tens of microns, all the architectures were successfully transferred to a large area substrate without any fracture within the structures, providing uniformity and scalability of transferring BTO/CFO structures using the stamping method.

**Discussion**

In summary, BTO/CFO free-standing thin films showed exceptionally high elastic strain and ME coupling. BTO/CFO films transferred to a PDMS substrate exhibited robust, repeatable and recoverable ME coupling behavior under high strains. Tunable ME coupling by stretching (88 % decrease in ME coefficient with 4.1 % of tensile strain) and its reversibility (cyclic stretch and release between 0 % and 3 % of strain) were demonstrated. It is possible to tailor the actual strain distribution in the materials and maintain constant ME coupling coefficient even under large stretching by adopting functional pattern designs such as fractals or auxetics. We believe that high-performance flexible ME composites and their capability to be transferred onto stretchable substrates with desired geometries will enable new advanced flexible electronics.

**Materials and Methods**

<u>Thin film deposition and fabrication</u>

BTO (15 nm)/CFO (15 nm) epitaxial thin films were deposited on (001) oriented MgO single crystalline substrate (Crystal GmbH) using pulsed laser deposition (Fig. S4) (*21*). A CFO layer was deposited at 550 °C and 10 mTorr oxygen partial pressure with the laser intensity and frequency of 1.8 J/cm$^2$ and 5 Hz, respectively. A BTO layer was then deposited at 750 °C and 200 mTorr oxygen partial pressure and the laser parameters of 1.2 J/cm$^2$ and 4 Hz. Subsequently, photoresist (AZ 1505) was spin-coated on the BTO/CFO thin films and patterned with different shapes using UV-photolithography. Patterned films were dry-etched with Ar-ion milling (Oxford IonFab 300 Plus) and the remaining photoresist was rinsed off with acetone (5 min), isopropyl alcohol (5 min), and oxygen plasma (600 W, 3 min). Afterwards, the MgO substrate was chemically etched with a sodium bicarbonate saturated solution and dried with a critical point dryer (Tousimis-CPD). As a result, various shapes of free-standing BTO/CFO structures were obtained (Fig. S5). Some of these free-standing structures were stamped with a polydimethylsiloxane (PDMS) substrate and were successfully transferred.

<u>Materials characterizations</u>

Piezoelectric properties and ME couplings were measured with piezoresponse force microscopy (PFM, ND-MDT) equipped with an in-plane DC magnetic field setup. A customized sample stage was designed in order to apply uniaxial strains on BTO/CFO//PDMS structures during the PFM measurements (Fig. S6). The BTO/CFO structures were patterned with bridge-like linear stripes of 1 μm width (Fig. S5) and transferred onto the PDMS substrate. During the PFM measurements, mechanical strain was applied vertically to the magnetic field direction (Fig. 3a). The strain was estimated by the elongated (stretched) width of the BTO/CFO stripes compared to the as-transferred width (Fig. S3). The ME coupling coefficients were calculated based on the local piezoelectric hysteresis loops averaged over five measurements.

In-situ nanomechanical tensile testing

For the quantitative investigation of the stretchability and the estimation of the maximum mechanical endurance without developing fractures like necks and cracks, the mechanical properties of BTO/CFO free-standing films, specifically failure stress and strain, were evaluated using nanomechanical force sensors. Tensile tests were performed using a scanning electron microscope (Nova NanoSEM 450, FEI company) equipped with a nanomechanical testing system (FT-NMT03, Femtotools AG) at room temperature. A BTO/CFO dogbone-like structure was attached to the tungsten force sensor probe (radius less than 0.1 µm) using SEM-compatible glue (SEMGLU, Kleindiek Nanotechnik GmbH) and the force was measured with 100 Hz sampling frequency under a 5 nm/s loading rate using a micro-electro-mechanical system (MEMS)-based force sensor (model FT-S200, ± 200 µN force range limit and 0.5 nN resolution). Inspired by the ASTM standard (22), we designed dogbone-like structures for mechanical tensile fracture tests. These structures were designed to have a narrow neck, where elongation and fractures in the material could take place. Two markers were added along the neck of the dogbone-like structure to analyze the elongation. The elongated lengths and strains of the dogbone-like structures were calculated based on the marker positions. Since the distance between the markers measured from the SEM images were highly dependent on the tilting of the sample, the geometrical configuration, particularly the tilting angle between the beam and the sample, was carefully controlled. Firstly, the initial distance between the two markers of the flat sample was measured (Fig. S7, left). The sample stage was then tilted and the sample was attached to the force sensor. To recover from the stage tilting, the attached sample was raised (Fig. S7, right).

**Acknowledgments**

This work has been financed by the ERC Consolidator Grant "Highly Integrated Nanoscale Robots for Targeted Delivery to the Central Nervous System" HINBOTS under grant no. 771565, the MSCA-ITN training program "mCBEEs" under grant no. 764977, the ERC Advanced Grant "Soft Micro Robotics" SOMBOT under grant no. 743217, and the Swiss National Science Foundation (Project No. 200021L_192012). X. C. would like to acknowledge the Swiss National Science Foundation (No. CRSK-2_190451) for partial financial support. M. K. acknowledges partial financial support from the Swiss National Science Foundation under Project No. 200021L_197017. H. C. acknowledges the National Research Foundation of Korea (2021M3F7A1082275 and 2017K1A1A2013237). J. P-L. acknowledges funding from the Swiss National Science Foundation, Project No. 200021_181988 and grant PID2020-116612RB-C33 funded by MCIN/ AEI /10.13039/501100011033. The authors would also like to thank the Scientific Center for Optical and Electron Microscopy (ScopeM), the FIRST laboratory at ETH for their technical support, and the Cleanroom Operations Team of the Binning and Rohrer Nanotechnology Center (BRNC) for their help and support.

**Funding:**
ERC Consolidator Grant "Highly Integrated Nanoscale Robots for Targeted Delivery to the Central Nervous System" (HINBOTS), grant no. 771565 (SP)
ERC Advanced Grant "Soft Micro Robotics" (SOMBOT), grant no. 743217 (BJN)

The MSCA-ITN training program (mCBEEs), grant no. 764977 (SP)
The Swiss National Science Foundation, Project No. 200021L_192012 (SP)
The Swiss National Science Foundation, Project No. CRSK-2_190451 (XZC)
The Swiss National Science Foundation, Project No. 200021L_197017 (BJN)
The National Research Foundation of Korea, Project No. 2021M3F7A1082275 (HC)
The National Research Foundation of Korea, Project No. 2017K1A1A2013237 (HC)
The Swiss National Science Foundation, Project No. 200021_181988 (JPL)
MCIN/AEI/10.13039/501100011033, grant PID2020-116612RB-C33 (JPL)

**Author contributions:**
    Conceptualization: MK, DK, XZC
    Methodology: MK, DK
    Investigation: MK, DK, XZC
    Data curation: MK, DK
    Formal analysis: MK, DK
    Visualization: MK, DK, XZC
    Funding acquisition: HC, JPL, BJN, XZC, SP
    Project administration: BJN, XZC, SP
    Supervision: XZC, SP
    Writing – original draft: MK, DK, HC, JPL, BJN, XZC, SP

**Data and materials availability:**

All data are available in the main text or the supplementary materials.

**Figures**

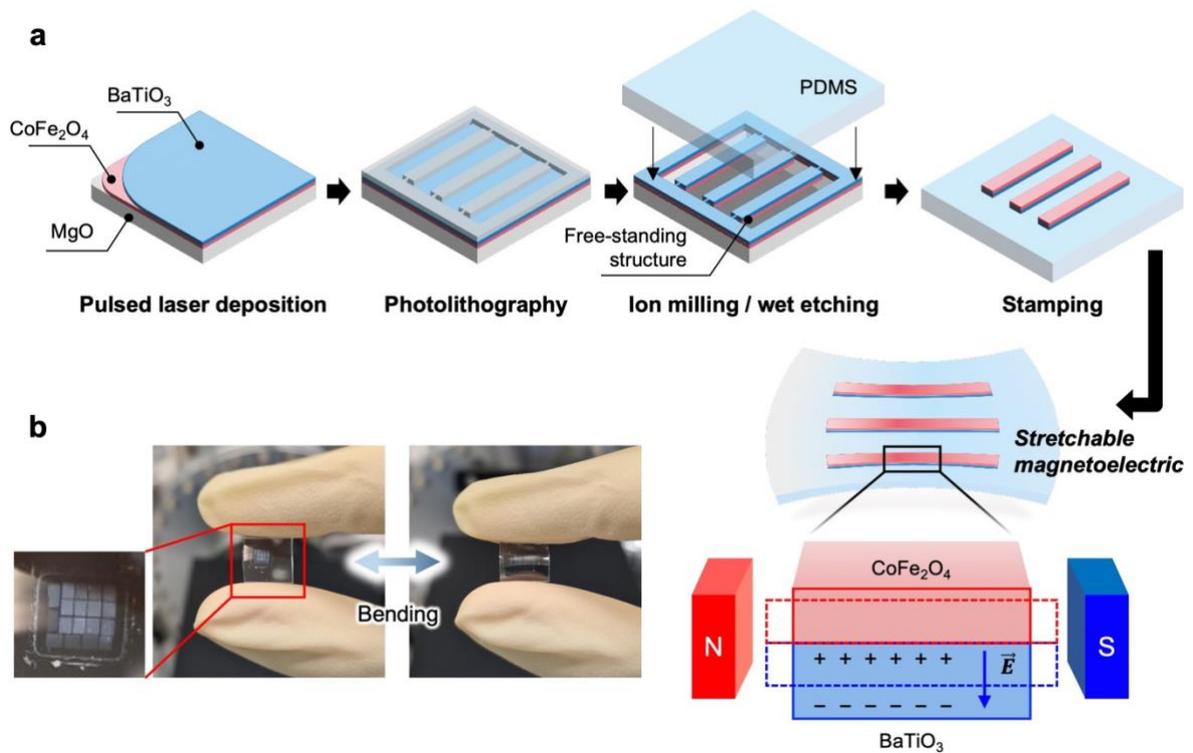

**Fig. 1. Flexible and stretchable magnetoelectrics.** (a) Schematic illustration of the fabrication processes of flexible and stretchable magnetoelectrics. Free-standing BTO/CFO structures were stamp-transferred onto the PDMS substrate. (b) Application of bending strains to the transferred BTO/CFO structures via PDMS substrate.

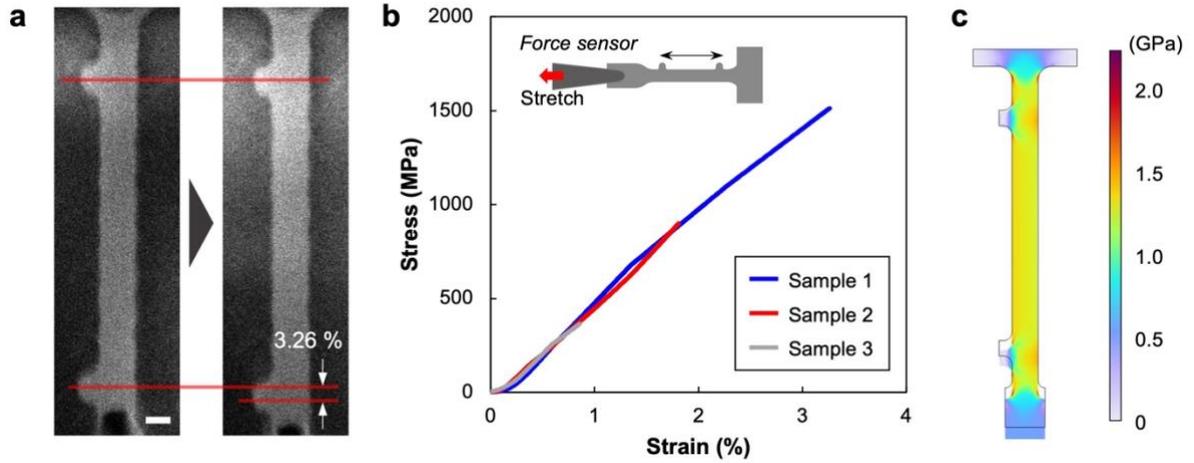

**Fig. 2. Mechanical tensile tests of BTO/CFO free-standing structures.** (a) SEM images of a dogbone-like structure during the tensile failure test. The scale bar indicates 1 µm. (b) Stress-strain curves obtained during the tensile failure test and the schematic illustration of the tensile test configuration (inset). (c) Finite element method-based mechanical simulations of dogbone-like structures.

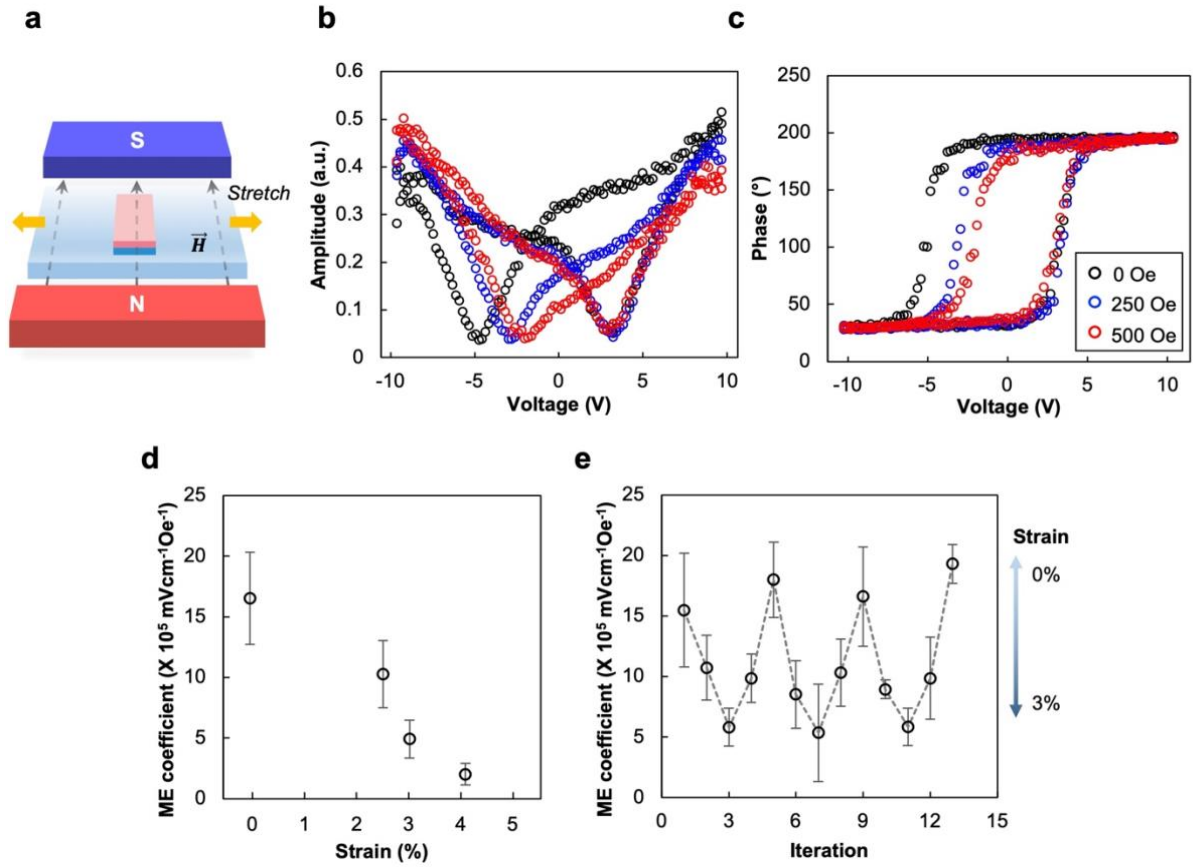

**Fig. 3. Reversible tuning of ME coupling.** (a) A schematic illustration of magnetic field application and mechanical stretching directions during PFM measurements. (b) Local PFM hysteresis amplitude and (c) phase of transferred BTO/CFO structures under magnetic fields. (d) ME coupling coefficients as a function of applied tensile strain. (e) Reversible tuning of ME coupling in transferred BTO/CFO structures.

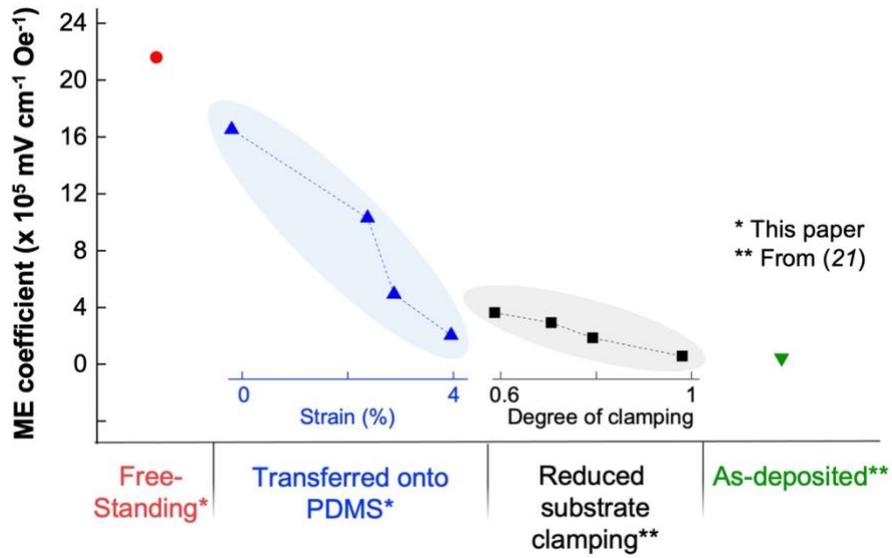

**Fig. 4. Comparison of the ME coupling coefficients.** ME coupling coefficients were calculated based on the PFM measurements on free-standing, transferred, patterned, and as-deposited BTO/CFO thin film structures. The data for the reduced substrate clamping and as-deposited thin films were taken from our previous work (*21*).

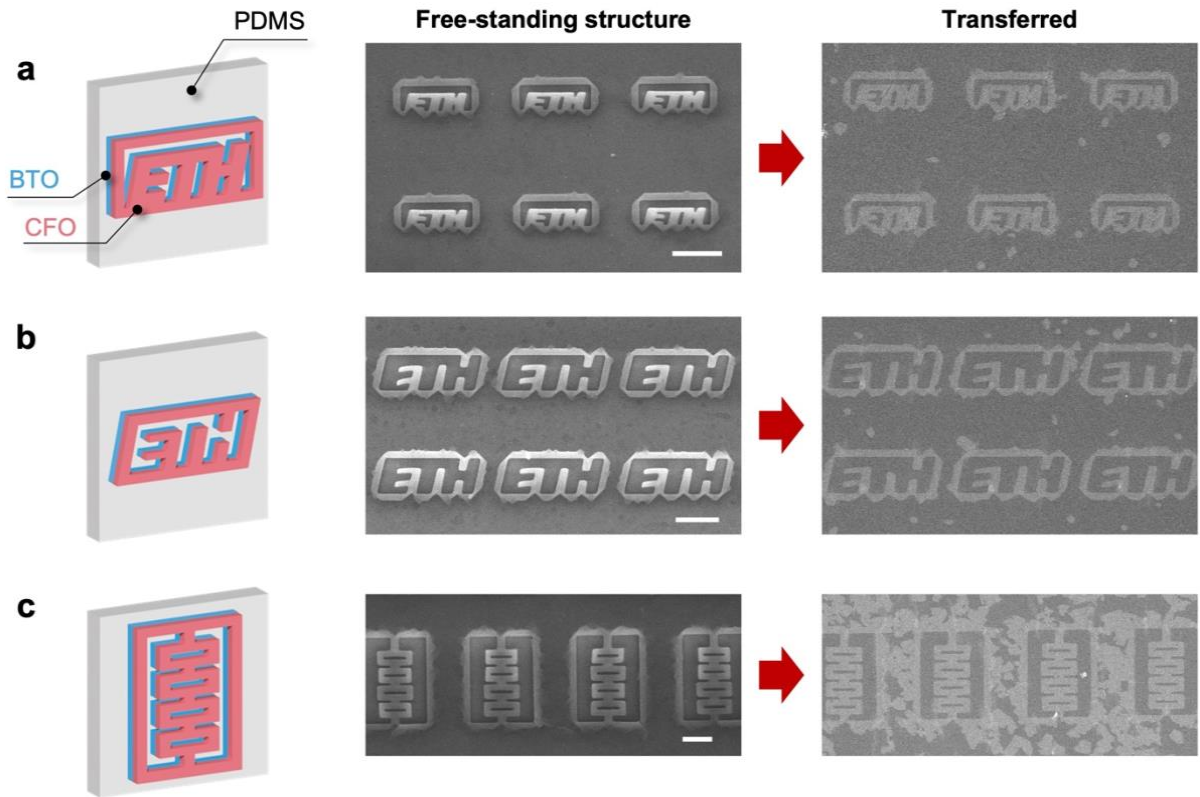

**Fig. 5. Transfer of various patterns of ME composite.** Stamp-transferring of various designs of BTO/CFO free-standing structures including (a) positively engraved ETH logo, (b) negatively engraved ETH logo, and (c) connected rectangular arrays.

# Supplementary Materials for

## Strain-sensitive flexible magnetoelectric ceramic nanocomposites

Minsoo Kim†, Donghoon Kim†, Buse Aktas, Hongsoo Choi, Josep Puigmartí-Luis, Bradley J. Nelson, Xiang-Zhong Chen*, Salvador Pané*

* Corresponding authors. Email: chenxian@ethz.ch, vidalp@ethz.ch

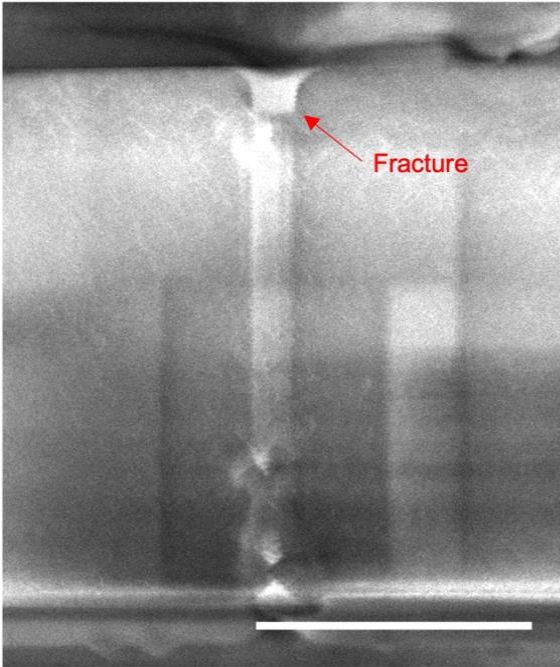 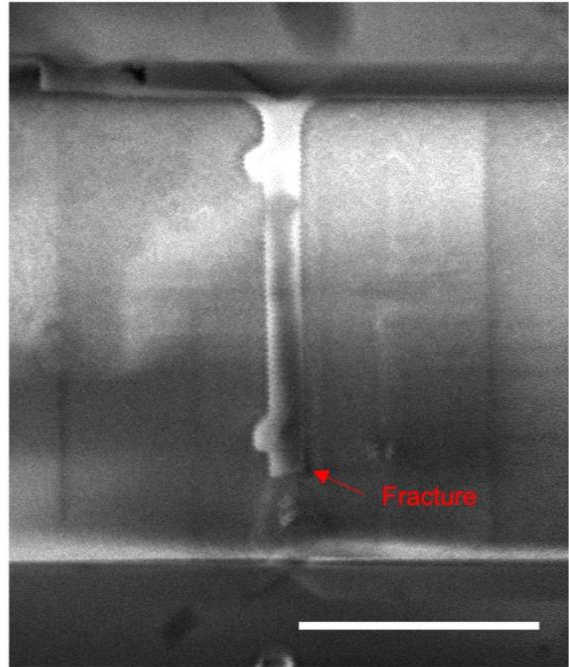

**Fig. S1.** Fracture point during the failure tests. Fracture happened near the marker of the BTO/CFO free-standing structures because of the stress concentration. Scale bars indicate 10 µm.

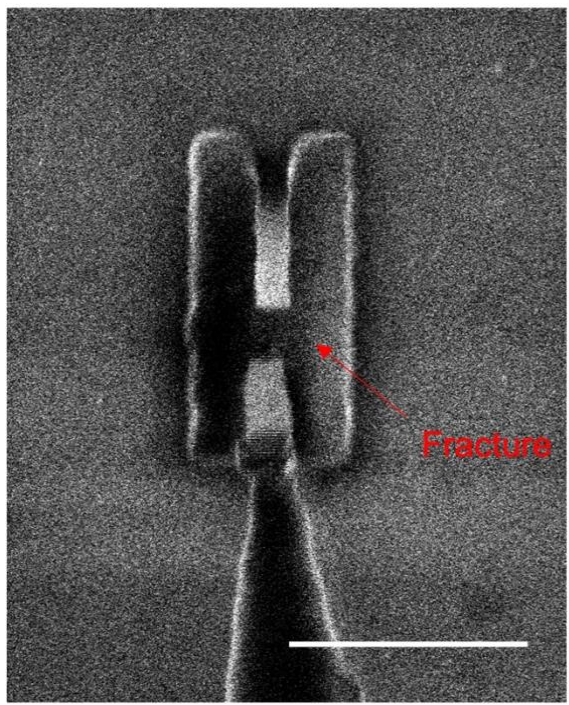 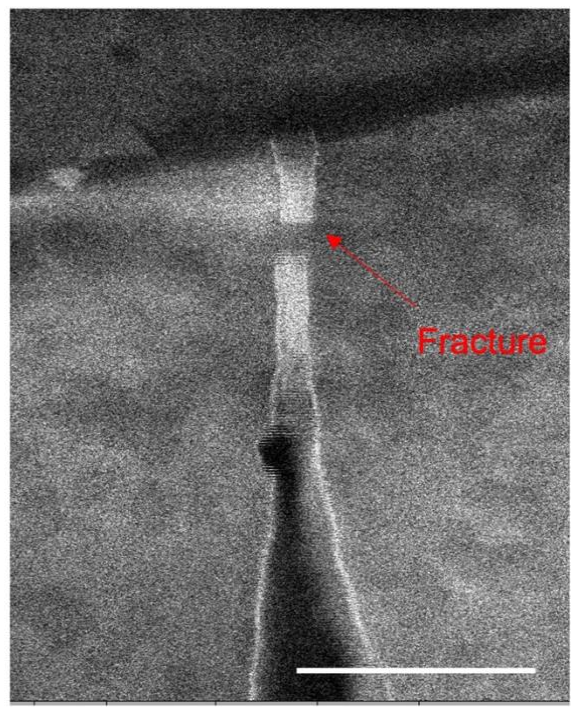

**Fig. S2.** Fractures happened in the neck for some cases of the dogbone-like structures without maker points. Scale bars indicate 10 µm.

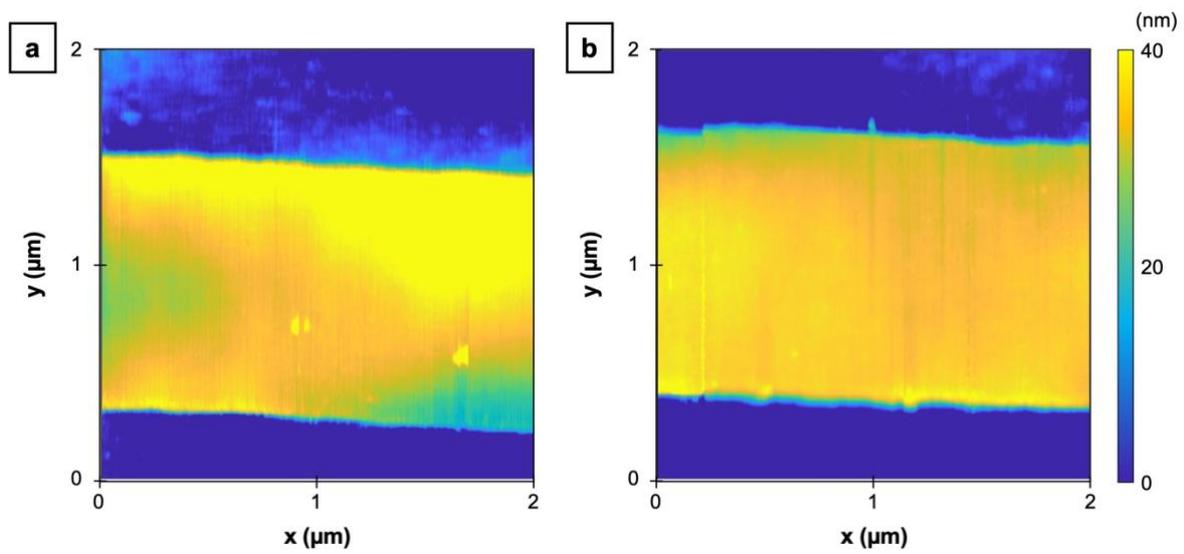

**Fig. S3.** AFM topography images of BTO/CFO linear stripes transferred onto the PDMS substrate. (a) BTO/CFO stripe before stretching PDMS substrate. (b) BTO/CFO stripe after stretching PDMS substrate. Based on the elongated width, the applied strain is calculated to 4.1 %.

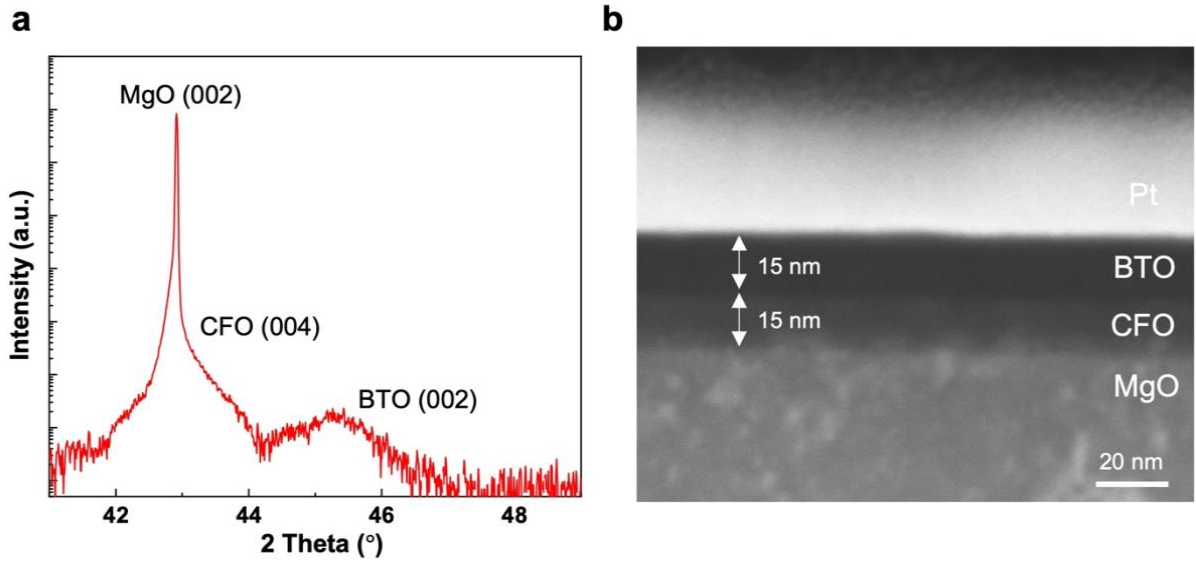

**Fig. S4.** (a) X-ray diffraction theta-2theta scan of as-deposited BTO/CFO//MgO (001) thin film. (b) Scanning electron microscope (SEM) image of the as-deposited BTO/CFO//MgO thin film.

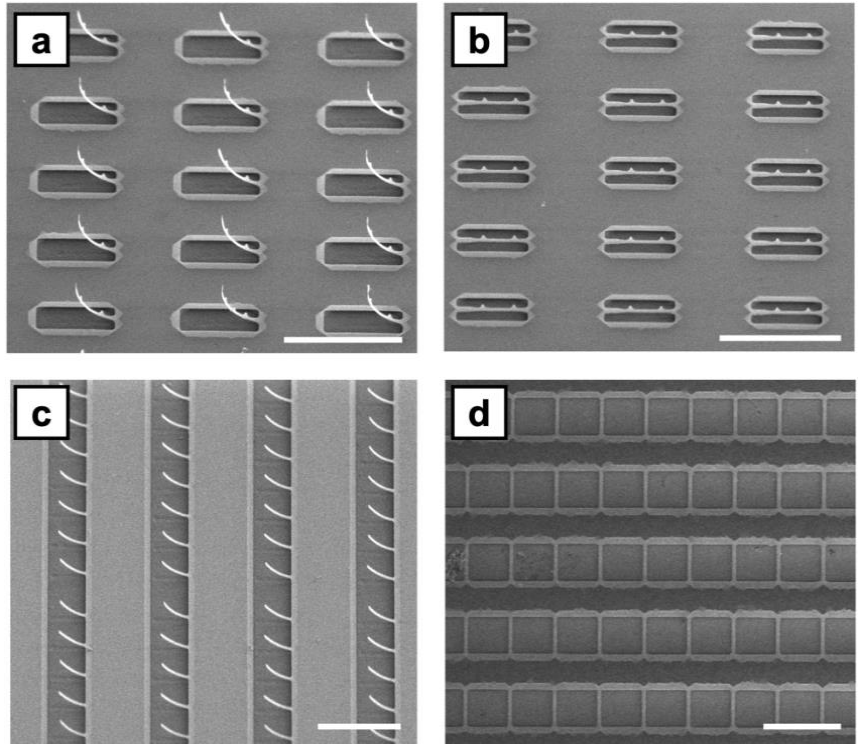

**Fig. S5.** Free-standing BTO/CFO structures after wet-etching of the MgO substrate. (a) One-side attached and (b) two-side attached SEM images of dogbone-like structures. (c) One-side attached and (d) two-side attached bridge-like linear stripes. All scale bars indicate 25 µm.

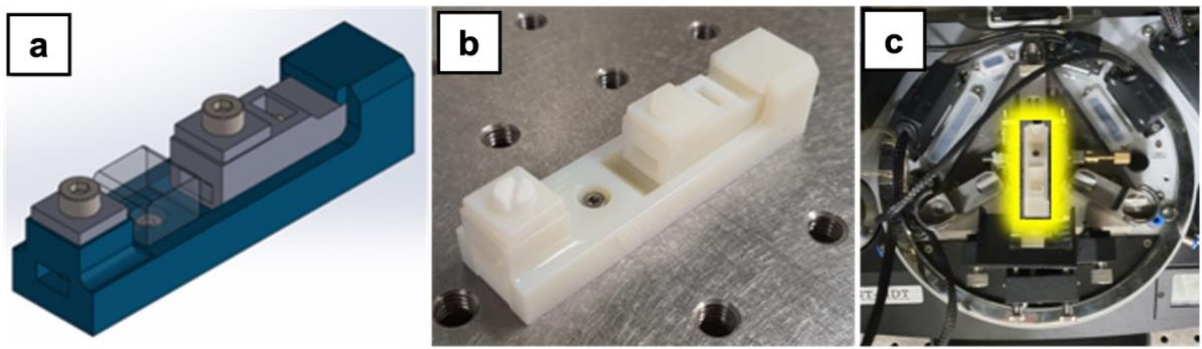

**Fig. S6.** Customized AFM-compatible sample holder for the *in-situ* tensile strain application. (a) The design and (b) 3-D printed structure of the sample holder. (c) The sample holder mounted on the AFM stage.

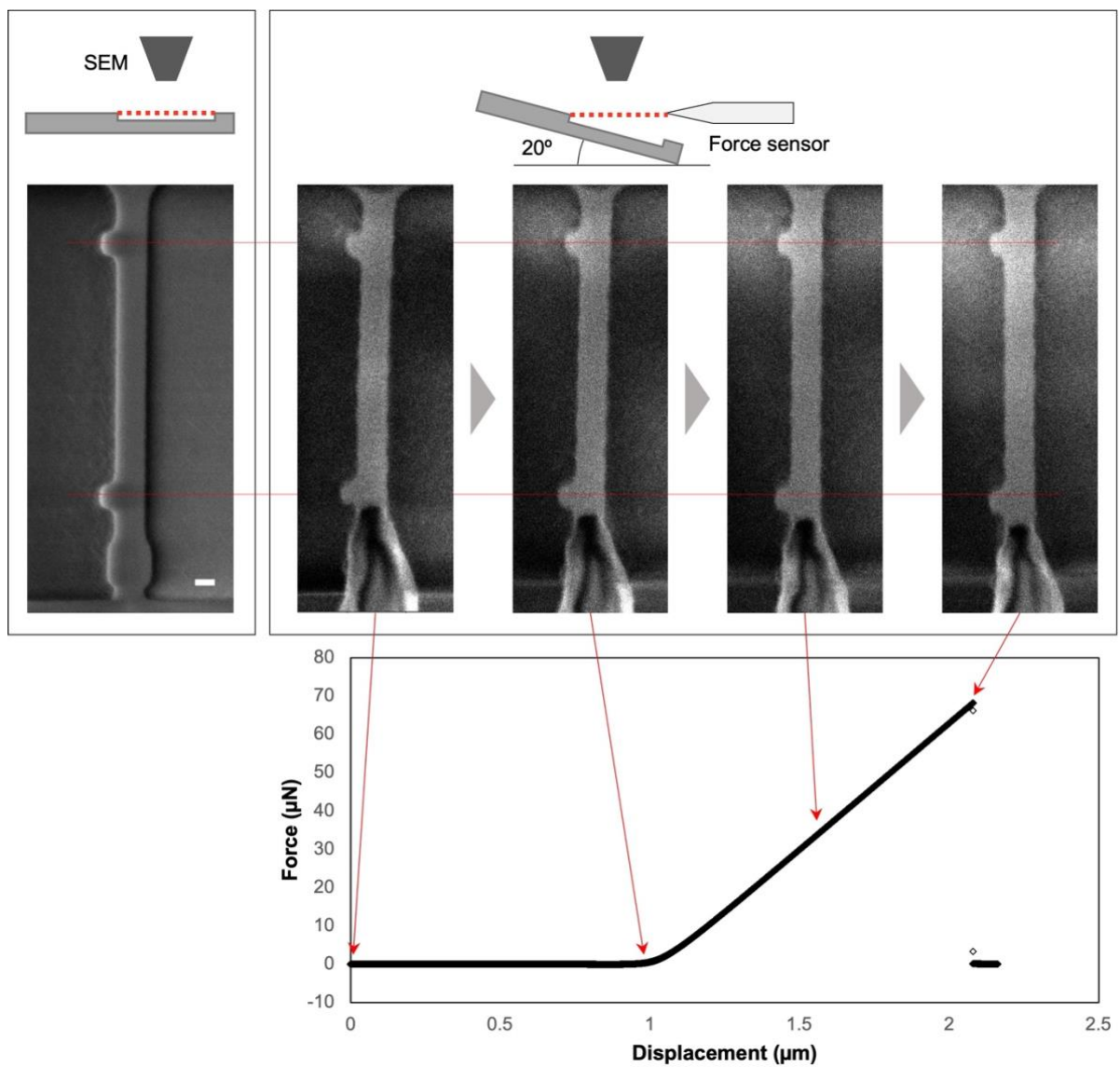

**Fig. S7.** Tensile failure test using a dogbone-like structure with obtained force-displacement data.

**Movie S1.**

Tensile failure test of a free-standing BTO/CFO dogbone-like structure and the force-displacement curve measured during the test.